\documentclass[11pt,a4paper]{article}
\usepackage{jheppub_kim}

\usepackage{pdflscape} 
\usepackage{amsmath}
\usepackage{amssymb}
\usepackage{dcolumn}
\usepackage{bm}
\usepackage{color}
\usepackage{epsfig}
\usepackage{amsfonts}
\usepackage{graphicx}
\usepackage{subfigure}
\usepackage{dcolumn}

\newcommand{\be}{\begin{equation}}
\newcommand{\ee}{\end{equation}}
\newcommand{\bea}{\begin{eqnarray}}
\newcommand{\eea}{\end{eqnarray}}
\newcommand{\nn}{\nonumber}

\def\be{\begin{equation}}
\def\ee{\end{equation}}
\def\bea{\begin{eqnarray}}
\def\eea{\end{eqnarray}}

\begin{document}

\title{Cosmological solutions and observational constraints on 5-dimensional braneworld
cosmology with gravitating Nambu-Goto matching conditions}

\author[a]{Georgios Kofinas}

\author[b,c]{Emmanuel N. Saridakis}

\author[d]{Jun-Qing Xia}

\affiliation[a]{Research Group of Geometry, Dynamical Systems and
Cosmology, Department of Information and Communication Systems Engineering,
University of the Aegean, Karlovassi 83200, Samos, Greece}

\affiliation[b]{Physics Division, National Technical University of Athens,
15780 Zografou Campus,  Athens, Greece}

\affiliation[c]{Instituto de F\'{\i}sica, Pontificia Universidad  Cat\'olica
de Valpara\'{\i}so, Casilla 4950, Valpara\'{\i}so, Chile}

\affiliation[d]{Key Laboratory of Particle Astrophysics, Institute of High
Energy Physics, Chinese Academy of Science, P.O. Box 918-3, Beijing 100049, China}

\emailAdd{gkofinas@aegean.gr}

\emailAdd{Emmanuel$_-$Saridakis@baylor.edu}

\emailAdd{xiajq@ihep.ac.cn}


\abstract{We investigate the cosmological implications of the
recently constructed 5-dimensional braneworld cosmology with
gravitating Nambu-Goto matching conditions. Inserting both matter
and radiation sectors, we extract analytical cosmological
solutions. Additionally, we use observational data from Type Ia
Supernovae (SNIa) and Baryon Acoustic Oscillations (BAO), along
with requirements of Big Bang Nucleosynthesis (BBN) and Cosmic Microwave
Background (CMB) radiation, in order to
impose constraints on the parameters of the model. We find that the
scenario at hand is in good agreement with observations, and
thus a small departure from the standard Randall-Sundrum scenario is
allowed. However, the concordance $\Lambda$CDM cosmology is still favored
comparing to both standard braneworld model and the present scenario.
}


\maketitle

\newpage

\section{Introduction}
\label{Introduction}

Israel matching conditions \cite{Israel 1966} are considered as the standard
equations of motion of a classical
codimension-1 defect which backreacts on the bulk dynamics. They are derived
by focusing on the
distributional part of the Einstein field equations (or some gravity
modification) where the brane
energy-momentum tensor, specified by a delta function, is included. An
equivalent way to derive these equations
is to take the variation of the brane-bulk action with respect to the induced
metric, while the bulk equations
of motion are derived as usually by varying the bulk action with respect to
the bulk metric. However, a
higher codimension defect carrying a generic energy-momentum tensor is known
to be inconsistent with Einstein's
equations \cite{Israel 1977,Geroch,Garfinkle:1999xv} (a brane with a pure
tension is a special consistent case
\cite{Vilenkin:1981zs,VilenkinPR,Vickers:1987az,Frolov:1989er,Unruh:1989hy,
Clarke:1990hu,
Nakamura:2003pi,Cline:2003ak,Apostolopoulos:2005ff}). In
\cite{Bostock:2003cv} it was considered the idea
that a more general theory like Einstein-Gauss-Bonnet gravity in six
dimensions might remove the previous inconsistency,
and the matching conditions of the theory for a generic energy-momentum
tensor were derived.
In \cite{CKP} the consistency of the whole system of bulk field equations
plus matching
conditions was shown for an axially symmetric codimension-2 cosmological
brane.

The spirit of the above proposal for consistency of the higher codimension
defects is to include higher Lovelock
densities \cite{Lovelock:1971yv,Zumino:1985dp}. However, in $D$ dimensions,
the highest such density is of
order $[(D-1)/2]$, and so, it is quite probable that branes with codimension
higher than $[(D-1)/2]$ will
still be inconsistent. Moreover, four-dimensions which represent effectively
spacetime at certain length
and energy scales do not allow generic codimension-2 or 3 defects. On the
other hand, Israel matching conditions and
their generalizations to higher codimensions do not accept the Nambu-Goto
probe limit, which is the lowest order
approximation of a test brane moving in curved background. Even the geodesic
limit of the Israel
matching conditions is questionable as a probe limit, since being the
geodesic equation a kinematical fact it should be
preserved independent of the gravitational theory (similarly to
\cite{GerochJang}, \cite{Ehlers:2003tv}) or the
codimension of the defect, which is not the case for these matching
conditions
\cite{Bostock:2003cv,Germani,Davis:2002gn,Gravanis:2002wy,Myers,Charmousis}.
Moreover, even the non-geodesic probe limit
of the standard equations of motion for various codimension defects in
Lovelock gravity theories is not
accepted, since this consists of higher order algebraic equations in the
extrinsic curvature, and therefore, a
multiplicity of probe solutions arise instead of a unique equation of motion
at the probe level. In view of these
observations a criticism to the standard matching conditions appeared in
\cite{kof-ira}, where alternative
matching conditions were proposed. These are the ``gravitating Nambu-Goto
matching conditions'' which arise
by the variation of the brane-bulk action with respect to the brane embedding
fields, so that the gravitational
back-reaction of the brane is taken into account. With these matching
conditions a brane is always consistent
for an arbitrary energy-momentum tensor and it also possesses the Nambu-Goto
probe limit (the codimension-2 case was
studied in \cite{kof-ira}, \cite{KofTom}, while the codimension-1 in
\cite{kof-zar}). In \cite{kof-zar} the application
of these alternative matching conditions led to a new 5-dimensional
braneworld cosmology which generalizes the conventional
braneworld cosmology \cite{binetruy} in the sense that it contains an extra
integration constant, and vanishing this
constant gives back the standard braneworld cosmology.

In the current work we try to confront this cosmology with the current
cosmological observational data (SNIa, BAO, BBN)
in order to construct the corresponding probability contour-plots for the
parameters of the theory. The  paper is
organized as follows: In section \ref{Introduction of the model} we briefly
present the alternative matching
conditions and the basic features behind these, and we find in the
cosmological
framework the equation for the expansion rate including both the matter and
radiation sectors. In section
\ref{Observational}, which is the main part of the work, we impose the
observational constraints on the parameters of
the model. Finally, a summary of the obtained results is given in section
\ref{Conclusions} of conclusions.

\section{5-dimensional braneworld with gravitating Nambu-Goto matching
conditions}

\label{Introduction of the model}

Our system is described by five-dimensional Einstein gravity coupled to a
localized 3-brane source.
The domain wall $\Sigma$ is assumed to be $Z_{2}-$symmetric, it splits the
spacetime $\mathcal{M}$ into two parts
$\mathcal{M}_{\pm}$ and the two sides of $\Sigma$ are denoted by
$\Sigma_{\pm}$. The total brane-bulk action is
\begin{eqnarray}
\!\!\!\!\!S= \int_{ \mathcal{M}} d^5x\sqrt{|\textsl{g}|}\,
\big(M^{3}\mathcal{R}-\Lambda\big)
-V \int_{ \Sigma}  d^4\chi\sqrt{|h|}-2M^{3} \int_{ \Sigma_{\pm}}
\!\!d^4\chi\sqrt{|h|}\,K
+\int_{ \Sigma} d^{4} \chi\,L_{mat}\,,\label{Stotal}
\end{eqnarray}
where $\textsl{g}_{\mu\nu}$ is the (continuous) bulk metric tensor and
$h_{\mu\nu}=\textsl{g}_{\mu\nu}-n_{\mu}n_{\nu}$ is the induced metric on the
brane with $n^{\mu}$ the unit normals
pointing inwards $\mathcal{M}_{\pm}$ ($\mu,\nu,...$ are five-dimensional
coordinate indices). The bulk coordinates are
$x^{\mu}$ and the brane coordinates are $\chi^{i}$ ($i,j,...$ are coordinate
indices on the brane). The brane tension
is $V>0$ and the matter Lagrangian of the brane is $L_{mat}$. The only matter
content of the bulk is the
cosmological constant $\Lambda < 0$ and the higher dimensional mass scale is
$M$. The contribution on each side of the
wall of the Gibbons-Hawking term is also necessary here as in the standard
derivation of the matching conditions.
$K=h^{\mu\nu}K_{\mu\nu}$ is the trace of the extrinsic curvature
$K_{\mu\nu}=h^{\kappa}_{\mu}h^{\lambda}_{\nu}n_{\kappa;\lambda}$ (the
covariant differentiation $;$ corresponds to
$\textsl{g}_{\mu\nu}$).

Varying (\ref{Stotal}) with respect to the bulk metric we get the bulk
equations of motion
\begin{equation}
\mathcal{G}_{\mu\nu} = -\frac{\Lambda}{2M^{3}}\textsl{g}_{\mu\nu}\,,
\label{bulk equations}
\end{equation}
where $\mathcal{G}_{\mu\nu}$ is the bulk Einstein tensor. In this variation,
beyond the basic
terms proportional to $\delta\textsl{g}_{\mu\nu}$ which give (\ref{bulk
equations}), there appear, as usually,
extra terms proportional to the second covariant derivatives
$(\delta\textsl{g}_{\mu\nu})_{;\kappa\lambda}$ which lead to a surface
integral on the brane with terms proportional to
$(\delta \textsl{g}_{\mu\nu})_{;\kappa}$. Adding the Gibbons-Hawking term,
the normal derivatives of
$\delta\textsl{g}_{\mu\nu}$, i.e. terms of the form
$n^{\kappa}(\delta\textsl{g}_{\mu\nu})_{;\kappa}$, are canceled, and
considering as boundary condition for the variation of the bulk metric its
vanishing on the brane (Dirichlet
boundary condition for $\delta\textsl{g}_{\mu\nu}$) there is nothing left
beyond the terms in
equation (\ref{bulk equations}). The Gibbons-Hawking term will again
contribute in the following variation
performed in order to obtain the brane equations of motion.
\par
According to the standard method, the interaction of the brane with the
bulk comes from the variation $\delta \textsl{g}_{\mu\nu}$ at the brane
position of the action (\ref{Stotal}),
which is equivalent to adding on the right-hand side of equation (\ref{bulk
equations}) the term $\kappa_{5}^2\,
\tilde{T}_{\mu\nu}\,\delta^{(1)}$, where
$\tilde{T}_{\mu\nu} =
\sqrt{|h|/|\textsl{g}|}\,\big(T_{\mu\nu}-\lambda\,h_{\mu\nu}\big)$,
$T_{\mu\nu}$ is the brane energy-momentum tensor and $\delta^{(1)}$ is the
one-dimensional delta function with support
on the defect. This approach leads to the standard Israel matching
conditions.
Here, we discuss an alternative approach where the interaction of the brane
with bulk gravity is obtained
by varying the total action (\ref{Stotal}) with respect to $\delta x^{\mu}$,
the embedding fields of the brane
position \cite{kof-ira}. The embedding fields are some functions
$x^{\mu}(\chi^{i})$ and their
variations are $\delta x^{\mu}(x^{\nu})$. While in the standard method the
variation of the bulk metric at
the brane position remains arbitrary, here the corresponding variation is
induced by $\delta x^{\mu}$, i.e.
$\delta \textsl{g}_{\mu \nu}=-\pounds_{\delta x} \textsl{g}_{\mu\nu}$.
The result of this variation gives the codimension-1 gravitating Nambu-Goto
matching conditions \cite{kof-zar}
(for a reminiscent variation see also \cite{Davidson})
\begin{eqnarray}
&&\Big[K^{ij}-Kh^{ij}+\frac{1}{4M^{3}}(T^{ij}-V h^{ij})\Big]
K_{ij}=0\label{match1Z}\\
&&T^{ij}_{\,\,\,\,|j}=-4M^{3}\big(K^{ij}-Kh^{ij}\big)_{|j}\,,
\label{match2Z}
\end{eqnarray}
where $K_{ij}=K_{ij}^{+}=K_{ij}^{-}$,
$K^{\mu\nu}=K^{ij}x^{\mu}_{\,\,,i}x^{\nu}_{\,\,,j}$ and $|$ denotes covariant
differentiation with respect to $h_{\mu\nu}$. These equations are
supplemented with the bulk equations
(\ref{bulk equations}) which are defined
limitingly on the brane, and therefore, additional equations have to be
satisfied at the brane position beyond the
matching conditions. Using these bulk equations the system of the above
matching conditions (\ref{match1Z}),
(\ref{match2Z}) is written equivalently as
\begin{eqnarray}
&&\big(T^{ij}-Vh^{ij}\big)K_{ij}=4(M^{3}R-\Lambda)
\label{cosmoequiv1Z}\\
&&T^{ij}_{\,\,\,\,|j}=0\,,
\label{cosmoequiv2Z}
\end{eqnarray}
where $R$ is the 3-dimensional Ricci scalar.

In order to search for cosmological solutions we consider the corresponding
form for the bulk metric in the
Gaussian-normal coordinates
\begin{equation}
ds_{5}^{2}=dy^{2}-n^{2}(t,y)dt^{2}+a^{2}(t,y)\,\gamma_{\hat{i}\hat{j}}(\chi^{
\hat{\ell}})
d\chi^{\hat{i}}d\chi^{\hat{j}}\,,
\label{bulk metric}
\end{equation}
where $\gamma_{\hat{i}\hat{j}}$ is a maximally symmetric 3-dimensional metric
($\hat{i},\hat{j},...=1,2,3$)
characterized by its spatial curvature $k=-1,0,1$. The energy-momentum tensor
on the brane $T_{ij}$ (beyond that of
the brane tension $V$) is assumed to be the one of perfect cosmic fluids with
total energy density $\rho$ and
total pressure $p$.

The $ty$, $yy$ bulk equations (\ref{bulk equations}) at the position of the
brane are
\begin{eqnarray}
&&\dot{A}+nH(A-N)=0\label{05cosmo}\\
&&A(A+N)-(X+Y)+\frac{\Lambda}{6M^{3}}=0\,,\label{55cosmo}
\end{eqnarray}
where
\begin{eqnarray}
&&A=\frac{a'}{a}\ ,\ \ N=\frac{n'}{n} ,\nonumber\\
&& H = \frac{\dot{a}}{na}\,,
\nonumber\\
&&X = H^{2} + \frac{k}{a^{2}}\,,\nonumber\\
&&Y = \frac{\dot{H}}{n} + H^{2}= \frac
{\dot{X}}{2nH} + X\,,
\label{notation2}
\end{eqnarray}
and a prime, a dot denote respectively $\partial/\partial{y}$,
$\partial/\partial{t}$.
The cosmic scale factor, lapse function and Hubble parameter arise as the
restrictions on the brane of the functions
$a(t,y),n(t,y)$ and $H(t,y)$ respectively. Other quantities also have their
corresponding values when
restricted on the brane, and since all the following equations will refer to
the brane position, we will use the same
symbols for the restricted quantities without confusion. Combining equations
(\ref{05cosmo}),
(\ref{55cosmo}) together with the matching condition (\ref{cosmoequiv1Z})
\cite{kof-zar}, we obtain the solution for
$A$
\begin{equation}
A=\pm\sqrt{X-\frac{\mathcal{C}}{a^{4}}-\frac{\Lambda}{12M^{3}}}\,\,,
\label{Abranches}
\end{equation}
where $\mathcal{C}$ is integration constant, and the Raychaudhuri equation
for the brane cosmology
\begin{eqnarray}
 \frac{\dot{H}}{n}+2H^{2}+\frac{k}{a^{2}}-\frac{
\Lambda}{6M^{3}} =\frac{\rho+3p-2V}{4M^{3}}\,
\frac{H^{2}+\frac{k}{a^{2}}-\frac{\mathcal{C}}{a^{4}}-\frac{\Lambda}{
12M^ { 3 } }}
{\frac{\rho+V}{4M^{3}}\pm6\sqrt{H^{2}+\frac{k}{a^{2}}-\frac{\mathcal{C}}{a^{4
}}
-\frac{\Lambda}{12M^{3}}}}\,.
\label{ram}
\end{eqnarray}
It is seen from (\ref{ram}) that for $\mathcal{C}=k=\rho=p=0$, the lower
branch contains the Minkowski
solution under the assumption of the Randall-Sundrum fine-tuning
$\Lambda+V^{2}/(12M^{3})=0$ \cite{rs1,rs2}. We will
not assume this condition in our analysis, so in the absence of matter our
cosmology may have a de-Sitter vacuum.
It is assumed that the quantity inside the square root of equation
(\ref{ram}) is positive.

In \cite{kof-zar} a single component perfect fluid was considered. Here,
since we want to confront the model
with real data, we will be more precise by assuming that the total energy
density $\rho$ consists of the matter
component $\rho_{m}$ with $p_{m}=0$ and the radiation component $\rho_{r}$
with $p_{r}=\frac{1}{3}\rho_{r}$,
i.e. $\rho=\rho_{m}+\rho_{r}$. Now, the integration process of (\ref{ram})
differs from that in \cite{kof-zar}.
The variable
\begin{equation}
\Xi=\frac{1}{2}\ln{\Big[\frac{12M^{3}}{-\Lambda}\Big(H^{2}+\frac{k}{a^{2}}
-\frac{\mathcal{C}}{a^{4}}
-\frac{\Lambda}{12M^{3}}\Big)\Big]}
\label{xi}
\end{equation}
obeys the differential equation
\begin{equation}
\frac{d\Xi}{d\ln{a}}=\frac{\tilde{\rho}+3\tilde{p}}{\tilde{\rho}\pm
6e^{\Xi}}-2\,, \label{xiequation}
\end{equation}
where
\begin{eqnarray}
&&\tilde{\rho}=\sqrt{\frac{12M^{3}}{-\Lambda}}\,\frac{\rho+V}{4M^{3}}=\frac{
\rho}{\rho_{\ast}}+\tilde{V}\label{1a}\\
&&\tilde{p}=\sqrt{\frac{12M^{3}}{-\Lambda}}\,\frac{p-V}{4M^{3}}=\frac{p}{
\rho_{\ast}}-\tilde{V}\label{2b}\\
&&\tilde{V}=\frac{V}{\rho_{\ast}}\label{3c}\\
&&
\rho_{\ast}=4M^{3}\sqrt{\frac{-\Lambda}{12M^{3}}}\,\,.\label{tilde}
\end{eqnarray}
Note that the Randall-Sundrum fine-tuning corresponds to the value
$\tilde{V}=3$.
Using the conservation equation (\ref{cosmoequiv2Z}) in the standard form
\begin{equation}
\dot{\rho}+3nH(\rho+p)=0\,,
\label{conserva}
\end{equation}
we obtain the equation
\begin{equation}
\frac{d\tilde{\rho}}{d\ln{a}}+3(\tilde{\rho}+\tilde{p})=0\,.
\label{conse}
\end{equation}
Finally, changing to the variable
\begin{equation}
\Phi=(\tilde{\rho}\pm 6 e^{\Xi})^{2}\,,
\label{Phi}
\end{equation}
we get from (\ref{xiequation}), (\ref{conse}), after some cancelations, the
differential equation
\begin{equation}
\frac{d\Phi}{d\ln{a}}+4\Phi=-2\tilde{\rho}(\tilde{\rho}+3\tilde{p})\,.
\label{ode}
\end{equation}
Each fluid component is conserved independently
\begin{equation}
\dot{\rho}_{m}+3nH(\rho_{m}+p_{m)}=0\quad,\quad
\dot{\rho}_{r}+3nH(\rho_{r}+p_{r})=0\,,
\label{conseindepe}
\end{equation}
so the solutions are
\begin{equation}
\rho_{m}=\frac{\rho_{m0}}{a^{3}}\quad,\quad\rho_{r}=\frac{\rho_{r0}}{a^{4}}
\,.
\label{sols}
\end{equation}
Therefore, equation (\ref{ode}) becomes a linear differential equation in
terms of $a$
\begin{equation}
\frac{d\Phi}{d\ln{a}}+4\Phi=-\frac{2}{\rho_{\ast}^{2}}\Big(\frac{\rho_{m0}}{
a^{3}}+\frac{\rho_{r0}}{a^{4}}+V\Big)
\Big(\frac{\rho_{m0}}{a^{3}}+2\frac{\rho_{r0}}{a^{4}}-2V\Big)\,,
\label{odas}
\end{equation}
with general solution
\begin{equation}
\Phi=\frac{1}{\rho_{\ast}^{2}}\big[\big(\rho_{m}+\rho_{r}+V\big)^{ 2}
-2V\rho_{r}\big]+\frac{\tilde{c}}{a^{4}}\,,
\label{soloPhi}
\end{equation}
where $\tilde{c}$ is integration constant.

From the definition (\ref{Phi}) we can find that
\begin{equation}
\tilde{\rho}\pm \frac{24M^{3}}{\rho_{\ast}}
\sqrt{H^{2}+\frac{k}{a^{2}}-\frac{\mathcal{C}}{a^{4}}-\frac{\Lambda}{12M^{3}}
}
=\epsilon\,\sqrt{\Phi}\,.
\label{sqrtPhi}
\end{equation}
In this equation the sign index $\epsilon=+1$ or $-1$ has been used to denote
a new different bifurcation
from the previous $\pm$ branches. It is seen from (\ref{sqrtPhi}) that the
sign $\epsilon=-1$ is only
consistent with the lower $\pm$ branch, while the sign $\epsilon=+1$ is
consistent with both $\pm$ branches.
The distinction, however, introduced by the sign index $\pm$ will be lost
in the expressions for the expansion rate and the acceleration parameter and
only the sign $\epsilon$
will distinguish the two branches of solutions.
\newline
The {\textit{expansion rate}} of the new cosmology arises by squaring
equation (\ref{sqrtPhi}) and is given by
\begin{equation}
H^{2}+\frac{k}{a^{2}}-\frac{\mathcal{C}}{a^{4}}=\Big(\frac{\rho_{\ast}}{
24M^{3}}\Big)^{ 2}\,
\Bigg{\{}\Bigg[\frac{\rho_{m} + \rho_{r}}{\rho_{\ast}}+\tilde{V} -\epsilon\
\sqrt{\Big(\frac{\rho_{m} + \rho_{r}}
{\rho_{\ast}}+\tilde{V}\Big)^{ 2}-2\tilde{V}\frac{\rho_{r}}{\rho_{\ast}}
+\frac{\tilde{c}}{a^{4}}}\,\Bigg]^{2} -36\Bigg{\}},
\label{H1b}
\end{equation}
where in (\ref{H1b}) one can set $\rho_{r}=0$. Redefining the integration
constant $\tilde{c}$ as
$\textsf{c}=\frac{\rho_{\ast}}{\rho_{r0}}\tilde{c}-2\tilde{V}$, the
expansion rate can also be written as
\begin{equation}
H^{2}+\frac{k}{a^{2}}-\frac{\mathcal{C}}{a^{4}}=\Big(\frac{\rho_{\ast}}{
24M^ {
3}}\Big)^{ 2}\,
\Bigg{\{}\Bigg[\frac{\rho_{m} + \rho_{r}}{\rho_{\ast}} + \tilde{V}
- \epsilon\,\sqrt{\Big(\frac{\rho_{m} + \rho_{r}}
{\rho_{\ast}} + \tilde{V}\Big)^{ 2} + \textsf{c}\frac{\rho_{r}}{\rho_{
\ast}}}\,\Bigg]^{2} -36\Bigg{\}},
\label{H}
\end{equation}
where in (\ref{H}) one cannot set $\rho_{r}=0$ since $\rho_{r0}$ is in the
denominator of the definition of
$\textsf{c}$. This solution contains two integrations constants. The first
constant $\mathcal{C}$ is associated
with the usual dark radiation term reflecting the non-vanishing bulk Weyl
tensor. The second constant
$\tilde{c}$ or $\textsf{c}$ is the new feature that does not appear in the
cosmology of the standard matching
conditions \cite{binetruy} and signals new characteristics in the cosmic
evolution. Setting $\textsf{c}=0\Leftrightarrow
\tilde{c}=\frac{2\tilde{V}\rho_{r0}}{\rho_{\ast}}$ in the
branch $\epsilon=-1$ we obtain the braneworld cosmology of the standard
matching conditions
$H^{2}+\frac{k}{a^{2}}-\frac{\mathcal{C}}{a^{4}}=
\big(\frac{\rho_{m}+\rho_{r}+V}{12M^{3}}\big)^{2}+\frac{\Lambda}{12M^{3}}$
(if there is no radiation
we just set $\tilde{c}=0$). Of course, there are always the extra
integration constants $\rho_{m0}$, $\rho_{r0}$ of equations (\ref{sols})
which are adjusted by
the today matter contents, while the today Hubble value $H_{0}$ is assumed to
be given.
The solution also contains three free parameters
$M$, $V$, $\Lambda$ or $M$, $\tilde{V}$, $\rho_{\ast}$.
In \cite{kof-zar} for a single dust perfect fluid, which approximates well at
least the late-times behaviour, it
was found analytically for values of $\tilde{V}$ extremely close to the
Randall-Sundrum fine-tuning the position
of the recent passage from a long deceleration era to the present
accelerating epoch. Moreover, the age of the
universe was estimated and the time variability of the dark energy
equation of state was calculated.

\section{Observational constraints}
\label{Observational}

As we analyzed in detail above, the cosmological scenario at hand leads to
the Friedmann equation (\ref{H1b}),
where the index $\epsilon=\pm1$ corresponds to two branches of solutions.
The Friedmann equation contains the following parameters:
$\mathcal{C}$,  $\tilde{c}$, $M$, $\tilde{V}$ and $\rho_{\ast}$, along with $\Omega_{m0}$, $\Omega_{r0}$,
$\Omega_{k0}$. $\mathcal{C}$ and $\tilde{c}$
are integration constants, $M$ is the fundamental 5D Planck mass,  and the other two $\tilde{V}$,
$\rho_{\ast}$ are connected to the fundamental model parameters $M$, $V$ and $\Lambda$ through the
relations (\ref{3c}), (\ref{tilde}). The identification of Newton's constant $G_{N}$ in equation
(\ref{H1b}) as a combination of the model parameters will reduce the number of these parameters by
one. Then, using $G_{N}$ we will define the various density parameters.

\subsection{Branch $\epsilon=-1$}
\label{epsilonminus1}

The scale factor for the branch $\epsilon=-1$ with $\tilde{V}< 3$ is bounded
from above and we will not consider this case in detail. However, the
branch $\epsilon=-1$ with
$\tilde{V}\geq 3$ possesses the late-times asymptotic linearized regime (that is when
$\rho_{m}+\rho_{r}<<\rho_{\ast}\tilde{V}$, $\rho_{r}/\rho_{r0}<<\tilde{V}^{2}/\tilde{c}$) with
a positive effective
cosmological constant
\begin{equation}
H^{2}+\frac{k}{a^{2}}\approx
\frac{\Lambda_{eff}}{3}+2\gamma\rho_{m}
+\gamma\rho_{r}+ \Big(\mathcal{C}+\frac{\gamma\rho_{\ast}\tilde{c}}{2\tilde{V}}\Big)\frac{1}{a^{4}}\,,
\label{linear1b}
\end{equation}
where
\begin{eqnarray}
&&\gamma=\frac{V}{144M^{6}}\label{gamma1}\\
&&\Lambda_{eff}=3{\Big({\frac{\rho_{\ast}}{4{M^3}
}}\Big)^{ 2}}\,\Big({\frac{\tilde{V}^2}{9}-1}\Big)
=\frac{1}{4M^3}\Big({\Lambda+\frac{V^2}{12 M^3}}\Big)\label{Lamdaeff}.
\end{eqnarray}
Now, as usual in braneworld or other modified gravity models, from this late-times Friedmann
equation, one reads the Newton's constant. Since asymptotically the coefficients of $\rho_{m},\rho_{r}$
in (\ref{linear1b}) are different, and
$\rho_{r} \ll \rho_{m}$, we associate Newton's constant with $\rho_{m}$
\begin{eqnarray}
&&\gamma=\frac{V}{144M^{6}}\equiv \frac{4\pi G_{N}}{3}\label{gamma}.
\end{eqnarray}
With this identification we can go back to the full Friedmann equation (\ref{H1b}) and
reduce one parameter, for instance $M$. Thus, the expansion rate (\ref{H1b}) for
$\epsilon=-1$, $\tilde{V}\geq 3$ becomes
\begin{equation}
H^{2}+\frac{k}{a^{2}}-\frac{\mathcal{C}}{a^{4}}=\frac{\pi
G_{N}\rho_{\ast}}{3\tilde{V}}\,
\Bigg\{\Bigg[\frac{\rho_{m} + \rho_{r}}{\rho_{\ast}}+\tilde{V}
+\sqrt{\Big(\frac{\rho_{m} + \rho_{r}}
{\rho_{\ast}}+\tilde{V}\Big)^{ 2}
-2\tilde{V}\frac{\rho_{r}}{\rho_{\ast}}+\frac{\tilde{c}}{a^4}}\,\Bigg]^{2} -36\Bigg\}\,.
\label{Hminus1b}
\end{equation}
Finally, in order to complete the steps we rewrite (\ref{Hminus1b}) as
\begin{equation}
H^{2}+\frac{k}{a^{2}}-\frac{\mathcal{C}}{a^{4}}= \frac{8\pi G_N}{3}
(\rho_m+\rho_r+\rho_{DE})
\label{Hminus1c}
\end{equation}
with
\begin{equation}
 \rho_{DE}= \frac{ \rho_{\ast}}{8\tilde{V}}\,
\Bigg\{\Bigg[\frac{\rho_{m} + \rho_{r}}{\rho_{\ast}}+\tilde{V}+\sqrt{
\Big(\frac{\rho_{m} + \rho_{r}}{\rho_{\ast}}+\tilde{V}\Big)^{2}
-2\tilde{V}\frac{\rho_{r}}{\rho_{\ast}}+\frac{\tilde{c}}{a^4}}
\,\Bigg]^{2} -36\Bigg\}-(\rho_m+\rho_r)\,.
\label{rhoDE1}
\end{equation}
Note that this $\rho_{DE}$ at late-times goes to $\frac{\Lambda_{eff}}{8\pi G_{N}}
-\frac{\rho_{r}}{2}+\frac{\rho_{\ast}\tilde{c}}{4\tilde{V}a^{4}}$ which asymptotically
goes to $\frac{\Lambda_{eff}}{8\pi G_{N}}$, i.e. to a simple cosmological constant.

So now, we can define the various density parameters straightforwardly as
\begin{eqnarray}
&& \Omega_m= \frac{8\pi G_N \rho_m}{3H^2}
\label{Omegam}\\
 &&\Omega_r= \frac{8\pi G_N \rho_r}{3H^2}
\label{Omegar}\\
 &&\Omega_{DE}= \frac{8\pi G_N \rho_{DE}}{3H^2}
\label{OmegaDE}\\
 &&\Omega_k= -\frac{k}{a^2H^2}
\label{Omegak}\\
 &&\Omega_{\mathcal{C}}= \frac{\mathcal{C}}{a^4H^2}.
\label{OmegaC}
\end{eqnarray}

Finally, assuming that the present scale factor is $a_0=1$ and using the
redshift as the independent variable ($1/a=1+z$), we can write the
Friedmann equation (\ref{Hminus1c}) in the usual form, convenient to observational fittings
\begin{equation}
\!\!\!H^2=H_0^2\left\{
\Omega_{k0}(1\!+\!z)^2+\Omega_{\mathcal{C}0}(1\!+\!z)^4+\Omega_{m0}(1\!+\!z)^3+
\Omega_{r0} (1\!+\!z)^4+ \frac{8\pi G_N \rho_{DE}(z)}{3H_0^2}\right\}\,.
\label{Hminus1dd}
\end{equation}
Here, $\rho_{DE}$, according to (\ref{rhoDE1}), is
\begin{eqnarray}
\rho_{DE}(z)&=&\frac{ \rho_{\ast}}{8\tilde{V}}\,
\Bigg\{\left[\frac{3H_0^2\Omega_{m0}}{8\pi
G_N \rho_{\ast}}(1+z)^3+\frac{3H_0^2\Omega_{r0}}{8\pi
G_N \rho_{\ast}}(1+z)^4+\tilde{V}+\mathcal{A}(z)
\right]^{2}-36\Bigg\}\nn\\
&&-\frac{3H_0^2\Omega_{m0}}{8\pi G_N}(1+z)^3-\frac{3H_0^2\Omega_{r0}}{8\pi
G_N}(1+z)^4\,,
\label{rhoDEz}
\end{eqnarray}
with
\begin{equation}
\mathcal{A}(z)=\sqrt{
\left(\frac{3H_0^2\Omega_{m0}}{8\pi
G_N \rho_{\ast}}(1\!+\!z)^3+\frac{3H_0^2\Omega_{r0}}{8\pi G_N \rho_{\ast}}(1\!+\!z)^4
+\tilde{V}\right)^{\!2}
-\frac{3H_0^2\Omega_{r0}\tilde{V}}{4\pi G_N \rho_{\ast}}(1\!+\!z)^4+\tilde{c}(1\!+\!z)^4}\,.
\end{equation}
Alternatively, one could write the last term inside the curly bracket of (\ref{Hminus1dd}) as
$\Omega_{DE0}(1+z)^{3(1+w_{DE}(z))}$, with
$\Omega_{DE0}=1-\Omega_{m0}-\Omega_{r0}-\Omega_{\mathcal{C}0}-\Omega_{k0}$
and $w_{DE}(z)$ extracted from (\ref{rhoDEz}). This normalization at the
current values fixes
one of the parameters, e.g. $\Omega_{r0}$.

In summary, Eq. (\ref{Hminus1dd}) is the one we will fit, with
$\mathcal{C}$,
$\tilde{c}$, $\tilde{V}$, $\rho_{\ast}$ and $\Omega_{m0}$ as parameters
(for simplicity we fix
$\Omega_{k0}$ to their Planck + WP + highL + BAO best fit values, namely
$\Omega_{k0}=-0.0003$
\cite{planckfit}). Concerning $H_0$ we include the direct $H_0$ probe from
the Hubble Space Telescope (HST) observations of Cepheid variables with
$H_0=73.8\pm2.4$ ${\rm km\,s^{-1}\,Mpc^{-1}}$, that is we set it as a free
parameter to fit the HST data.
\begin{figure}[ht]
\begin{center}
\includegraphics[width=10cm]{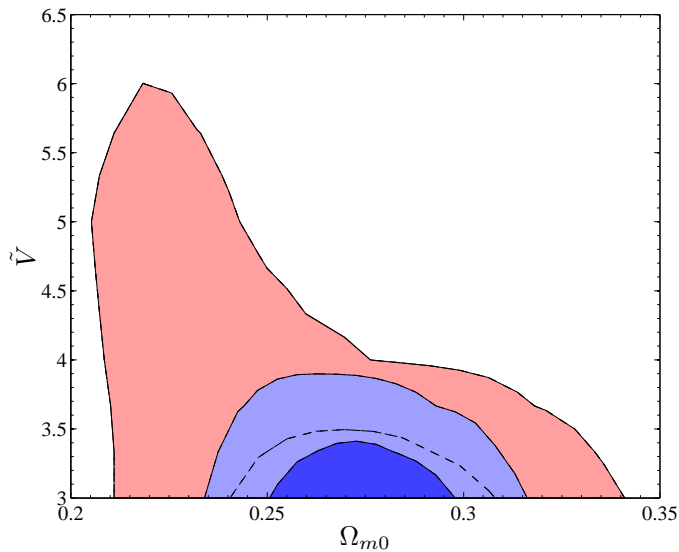}
\caption{(Color Online) {\it{Two-dimensional  likelihood contours in the
($\Omega_{m0},\tilde{V}$) plane for the $\epsilon=-1$ branch and fixed
$\tilde{c}$
to its Randall-Sundrum value ($\tilde{c}=2\tilde{V}\rho_{r0}/\rho_\ast$)
from the SnIa (red and pink) and SnIa+BAO (blue and light blue) data
combinations. The light regions (pink and light blue respectively) correspond
to 2$\sigma$ confidence level, while the darker regions (red and blue
respectively) correspond to 1$\sigma$ confidence
level. Note that in this specific plot the 1$\sigma$ bound of the  SnIa (red)
data combinations is inside the 2$\sigma$ bound of the  SnIa+BAO (light blue)
data combinations.
}}} \label{omm_v}
\end{center}
\end{figure}

The $\mathcal{C}$-term in (\ref{Hminus1c}) corresponds to
dark radiation, so it is proportional to $1/a^4$. This term, in particular
$\Omega_{\mathcal{C}0}$, cannot be constrained efficiently by the
low-redshift observations we are going to use in our analysis. However,
since this dark radiation component was present at the time of
Big Bang Nucleosynthesis (BBN) too, that is at redshift $z_{BBN}\sim 10^9$,
we can use BBN arguments in order to constrain it. Specifically, the data
impose an upper bound on the amount of total radiation (standard and
exotic), which is expressed through the parameter $\Delta N_\nu$ of the
effective neutrino species \cite{bbn,Dutta:2009jn,Dutta:2010jh,Ichiki:2002eh}. Thus, in our case, this
bound imposes a constraint on $\Omega_{\mathcal{C}0}$, namely
\begin{equation}
\Omega_{\mathcal{C}0} = 0.135 \Delta N_\nu \Omega_{r0}~.
\end{equation}
The recently released Planck results impose a quite tight constraint
on the effective number of neutrino species \cite{planckfit}: $N_{\rm eff} =
3.30^{+0.54}_{-0.51}$ (95\% C.L.) from the Planck+WP+highL+BAO data
combination. Therefore, the 95\% C.L. upper limit of $\Delta N_\nu$ is
$\Delta N_\nu < 0.776$. This leads to a very tight constraint on   the dark
radiation component of the scenario at hand, namely $\Omega_{\mathcal{C}0} <
5\times10^{-6}$ (95\% C.L.). Thus, we can safely neglect this term in the
remaining analysis and the remaining parameters to be fitted are
$\tilde{c}$, $\tilde{V}$, $\rho_{\ast}$ and $\Omega_{m0}$.
\begin{figure}[ht]
\begin{center}
\includegraphics[width=9cm]{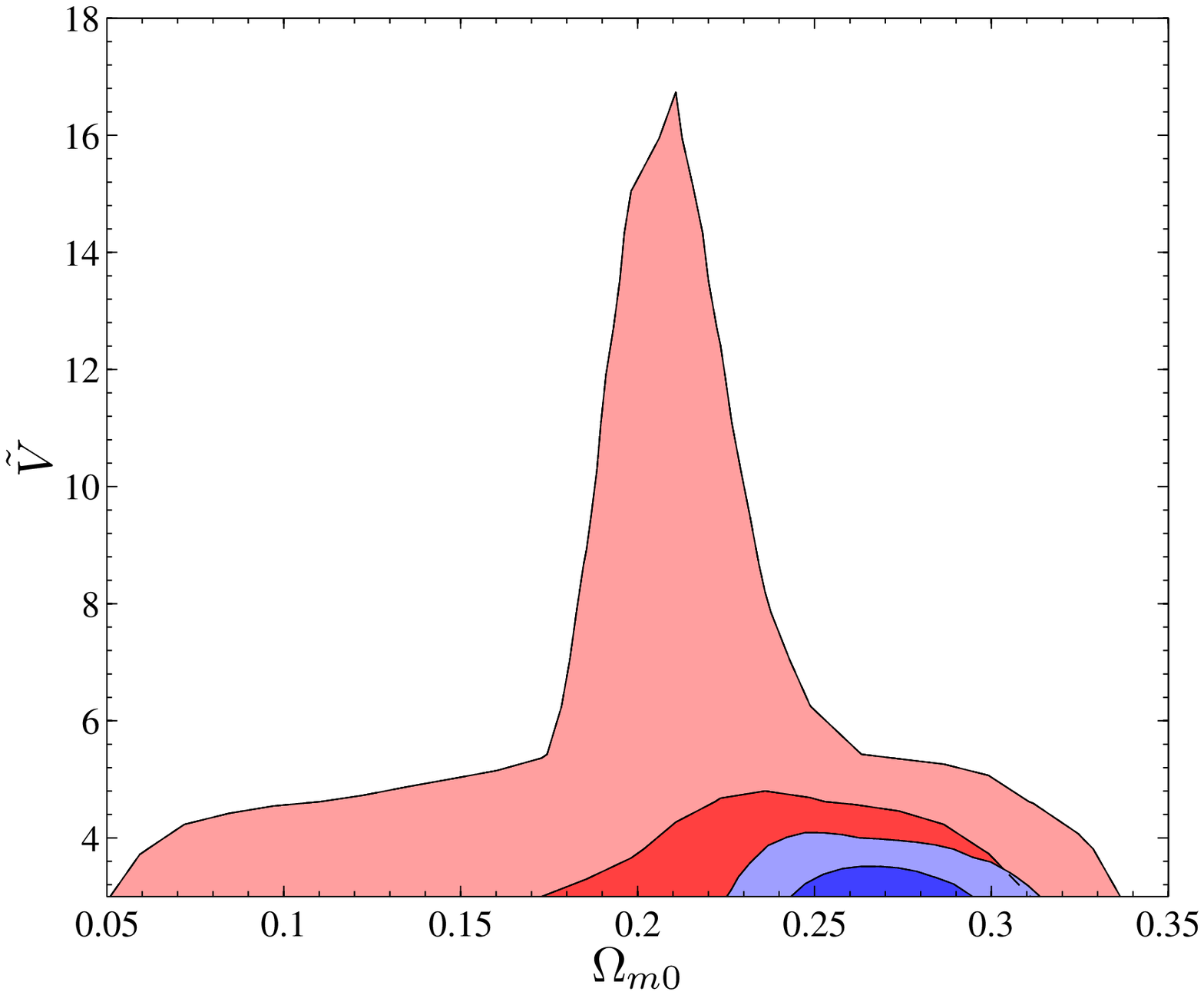}
\includegraphics[width=9cm]{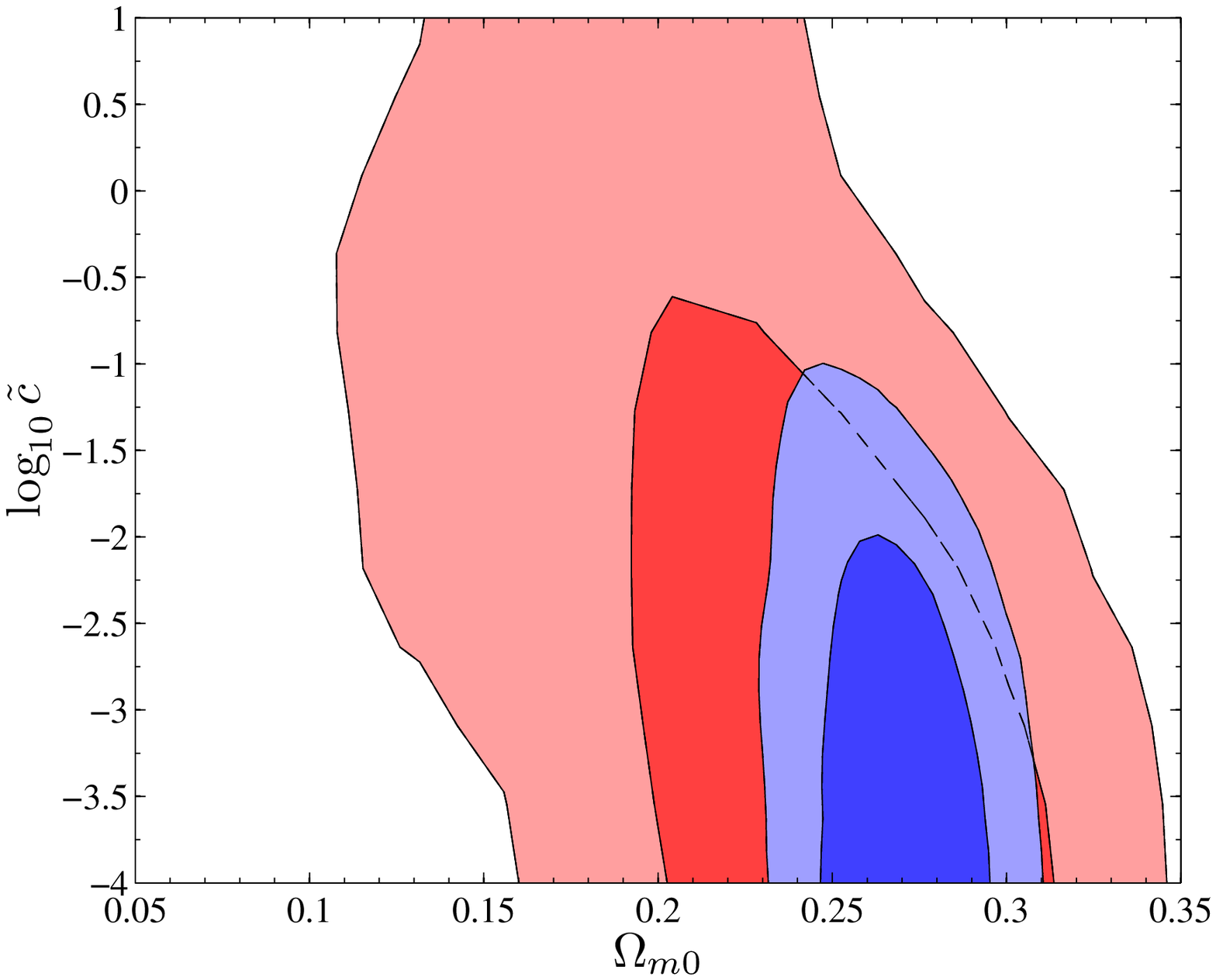}
\caption{(Color Online) {\it{Two-dimensional likelihood contours in the
($\Omega_{m0},\tilde{V}$) and ($\Omega_{m0},\log_{10}\tilde{c}$) planes for
the $\epsilon=-1$ branch from the SnIa (red and pink) and SnIa+BAO (blue and
light blue) data combinations. The light regions (pink and light blue
respectively)
correspond to 2$\sigma$ confidence level, while the darker regions (red and
blue
respectively) correspond to 1$\sigma$ confidence level.}}} \label{omm_v2}
\end{center}
\end{figure}

As a starting analysis, let us fit the case where $\tilde{c}$ is set to its
value that corresponds to the standard braneworld cosmological scenario \cite{binetruy}, namely
$\tilde{c}=2\tilde{V}\rho_{r0}/\rho_\ast$ (which is exactly zero in the
absence of radiation). Thus, in this
case we have only three free parameters, namely $\tilde{V}$, $\rho_{\ast}$ and $\Omega_{m0}$.
In Fig. \ref{omm_v} we provide the two-dimensional contour plots on
($\Omega_{m0},\tilde{V}$), using SnIa and SnIa+BAO data combinations. The details
of the fitting procedure are presented in the Appendix. As we observe, when
we use SnIa data only, the constraints on $\tilde{V}$ are relatively weak,
namely $3<\tilde{V}<5.5$ at the $95\%$ confidence level. However, addition
of the BAO data introduces an extra constraining power and the total
constraint becomes tighter, namely $3<\tilde{V}<3.4$ ($95\%$ C.L.) from
SnIa+BAO data. Finally, as we describe in the Appendix, the efficiency of the
fitting is quantified by $\chi^2$, which for this case is
$\chi^2\approx570$.

Let us now proceed to the general case, that is considering  $\tilde{c}$ as
an additional free parameter. In the upper graph of Fig. \ref{omm_v2} we
present the contour plots  of $\tilde{V}$ versus $\Omega_{m0}$, while in the
lower graph of Fig. \ref{omm_v2} we depict the contour plots of $\tilde{c}$
versus $\Omega_{m0}$. As we observe, the  SnIa constraints on
the parameter $\tilde{V}$ are much weaker than those of Fig. \ref{omm_v},
due to the additional fitting variable. In particular,
the 95\% C.L. bound is $3<\tilde{V}<15.3$ (additionally note that   the
parameter space $\Omega_{m0} < 0.2$ is now allowed by the SnIa data, exactly
due
to the presence of non-zero $\tilde{c}$). Concerning $\tilde{c}$ the  SnIa
data leads also to the relatively weak constraint $\log_{10}\tilde{c} < 0.1$
(95\% C.L.). However, for the combined SnIa with BAO data, the constraints become
much tighter. At 95\% confidence level they are
$3<\tilde{V}<3.7$ and $\log_{10}\tilde{c} < -1.6$, while their best fit
values are very close to 3 and 0 respectively. Finally, the
corresponding
$\chi^2$ is $\chi^2\approx570$.
\begin{figure}[ht]
\begin{center}
\includegraphics[width=10cm]{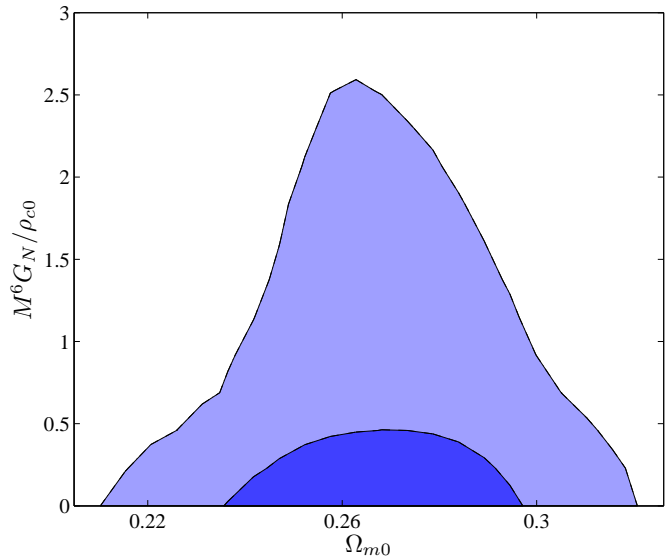}
\caption{(Color Online) {\it{Two-dimensional likelihood contours of the
dimensionless quantity $M^{6}G_{N}/\rho_{c0}$ versus $\Omega_{m0}$, where
$\rho_{c0}$ is the current critical density,
for the $\epsilon=-1$ branch from the SnIa+BAO data
combinations. The lighter region corresponds
to 2$\sigma$ confidence level, while the darker region  corresponds to
1$\sigma$ confidence level.}}} \label{massM}
\end{center}
\end{figure}

Let us now refer to the constraints of the Cosmic Microwave Background (CMB)
radiation on the scenario at hand. One can use such high-redshift probes, in
particular the distance information of CMB, the shift parameter $R$ and the
acoustic scale $\ell_A$, from WMAP9 or Planck data \cite{planckfit}. Their
definitions are
$R=\sqrt{\Omega_mH^2_0}\chi(z_\ast)/c$ and
$\ell_A=\pi\chi(z_\ast)/\chi_s(z_\ast)$, where $\chi(z_\ast)$ and
$\chi_s(z_\ast)$ denote the comoving distance to the decoupling epoch
$z_\ast$ and the comoving sound horizon at $z_\ast$, respectively.
The current CMB data imply that at $z_\ast \sim 1100$ the Universe
is dominated by matter, that is $H(z)^2 \sim \rho_m(z)$. If in our model we
neglect $\Omega_k$, $\Omega_r$ and $\Omega_{C}$ terms,
and we insert the present value of the critical density
 $\rho_{c0}={3H_0^2}/{8\pi G_N}$,
then  (\ref{Hminus1c}) becomes
$H^2=\rho_m(z)+\rho_{DE}(z)=H_0^2\Omega(z)$, where
\begin{eqnarray}
\Omega(z)
&=& \frac{1}{2}\left\{\left[\frac{{\Omega_{m0}}(1+z)^3}{\sqrt{
\Omega_\ast\tilde{V}}}+\sqrt{\Omega_\ast\tilde{V}}\right]^{2}
-9\frac{\Omega_\ast}{\tilde{V}}\right\},
\end{eqnarray}
with
\begin{equation}
\Omega_\ast \equiv\frac{\rho_\ast}{\rho_{c0}}.
\label{Omegastar}
\end{equation}
Apparently, we deduce that if we want the term $\rho_m(z)^2$ to be
significantly smaller than $\rho_m(z)$, we need
\begin{eqnarray}
\frac{\left[\Omega_m(1+z)^3\right]^2}{\Omega_\ast\tilde{V}} \ll 2
(\Omega_m(1+z)^3)\ \
\Rightarrow\ \  \frac{\Omega_m(1+z)^3}{2\tilde{V}}\ll \Omega_\ast~.
\end{eqnarray}
Since $\tilde{V}\gtrsim3$, it is implied that if we desire to satisfy the
CMB data we need $\Omega_\ast \gg 0.05(1+z_\ast)^3\sim10^7$.

Proceeding forward, combining equations (\ref{3c}), (\ref{gamma}) we obtain
for the fundamental
mass  scale $M$ the relation
\begin{equation}
M^{6}=\frac{\tilde{V}\rho_{\ast}}{192\pi G_{N}}\,.
\label{Mmplanck}
\end{equation}
The likelihood contours of the dimensionless quantity  $M^{6}G_{N}/\rho_{c0}$
versus $\Omega_{m0}$ is
shown in Fig. \ref{massM}. We can then straightforwardly
estimate that at 1$\sigma$ confidence level
$0 <M<0.042\,\text{GeV}$. Moreover, to give an estimate for  the value of the
brane tension $V$, we use  the relation $V=192\pi G_N M^6$, which leads to $0
<V<2.22\times 10^{-44}\,\text{GeV}^4$ at 1$\sigma$ confidence level. That is
$0<V<0.87\times 10^{3}\rho_{\Lambda0}$, where $\rho_{\Lambda0}$ is the
current value of the energy density of the observed cosmological constant.

\begin{figure}[ht]
\begin{center}
\includegraphics[width=10cm]{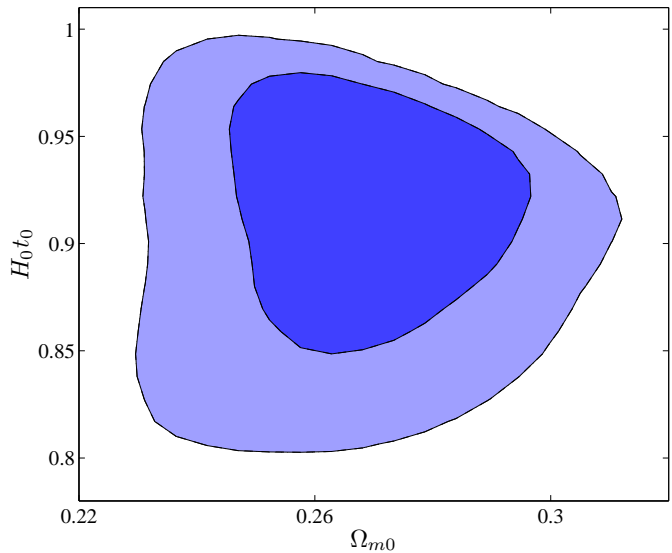}
\caption{(Color Online) {\it{Two-dimensional likelihood contours of $H_0
t_0$ versus $\Omega_{m0}$
for the $\epsilon=-1$ branch from the SnIa+BAO data
combinations. The lighter region corresponds
to 2$\sigma$ confidence level, while the darker region  corresponds to
1$\sigma$ confidence level.}}} \label{agee}
\end{center}
\end{figure}
Finally, we close this subsection by examining the constraints on the model
from the age of the universe. In general, the age of the universe is given by
\begin{equation}
t_{0}=\int_{0}^{\infty}\frac{dz}{(1+z)H(z)}\,,
\label{age}
\end{equation}
where in the scenario at hand $H(z)$ is given by equation (\ref{Hminus1dd}).
Thus, taking into account the constraints on the model parameters elaborated
above, we can construct the contour plots of $H_0 t_0$ versus $\Omega_{m0}$,
which is presented in Fig. \ref{agee}. We can then straightforwardly
estimate the age in Gyr, finding  $12.23\ \text{Gyr}\leq
t_0\leq14.13\ \text{Gyr}$ at 1$\sigma$ confidence level (for the
$\Lambda$CDM model with $\Omega_{m0}=0.28$ the corresponding age is
13.5$\,$Gyr). We observe from equations (\ref{H1b}) and (\ref{age}) that
larger values of the mass scale $M$ in the range found above correspond to
larger values of the age of the universe. Thus, since larger ages are
preferable, the most probable estimations for $M$ lie closer to the upper
bound.

In summary, as we observe, the cosmological observations constrain
$\tilde{V}$ and
$\tilde{c}$ close to their  Randall-Sundrum values, namely $\tilde{V}=3$ and
$\tilde{c}\approx0$ ($\tilde{c}=0$ in the case of radiation absence).
However, note that the data allow for a departure from Randall-Sundrum
scenario. In particular,
although the present model has an additional parameter compared to
Randall-Sundrum one, the corresponding $\chi^2$ is the same in two models,
namely $\chi^2\approx570$. This means that braneworld models with gravitating
Nambu-Goto matching condition are in ``equal'' agreement with observations
as the standard braneworld models.

Lastly, if we desire to compare the scenario at hand with the concordance
paradigm of standard $\Lambda$CDM cosmology, we can be based on the Akaike
Information Criterion (AIC)
\cite{Akaike74}
\begin{equation}
AIC = -2\ln{\mathcal{L}_{max}} + 2k\,,
\label{AIC}
\end{equation}
where
$\,
\ln{\mathcal{L}_{max}} = -\chi^2_{min} / 2
$
is the maximum likelihood achievable by the
model (with $\chi^2_{min} / 2$ the corresponding $\chi^2$ of the
analysis) and $k$ the number of parameters of the model. Hence, we obtain the
difference on the AIC between the standard
$\Lambda$CDM cosmology and our gravitating Nambu-Goto matching conditions model as
\begin{eqnarray}
&&\!\!\!\!\!\!\!\!\!\!\!\Delta AIC = AIC (\text{gravitating Nambu-Goto match.
cond.}) - AIC
(\Lambda\text{CDM})\nonumber\\
&&\ \ \ \ =\chi^2_{min} (\text{gravitating Nambu-Goto match. cond.}) -
\chi^2_{min} (\Lambda\text{CDM}) + 2\Delta{k},
\end{eqnarray}
where $\Delta{k} = k (\text{gravitating Nambu-Goto match.
cond.}) - k(\Lambda\text{CDM})$ is the difference of the number
of parameters between the models. Thus, although in our model we obtain a
$\chi^2_{min}$ similar to that of  $\Lambda$CDM
cosmology ($\chi^2_{min} (\Lambda\text{CDM})\approx570$), the fact that we
use two
additional parameters gives $\Delta AIC \approx 4$. Thus, we deduce that
$\Lambda$CDM cosmology is more favored comparing to
the scenario at hand, since the two extra parameters do not improve the
late-times fitting behavior.

\subsection{Branch $\epsilon=+1$}

In this case, the full Friedmann equation (\ref{H1b}) is
\begin{equation}
H^{2}+\frac{k}{a^{2}}-\frac{\mathcal{C}}{a^{4}}=\Big(\frac{\rho_{\ast}}{
24M^{3}}\Big)^{\!2}\,
\Bigg{\{}\Bigg[\frac{\rho_{m} + \rho_{r}}{\rho_{\ast}}+\tilde{V} -\
\sqrt{\Big(\frac{\rho_{m} + \rho_{r}}
{\rho_{\ast}}+\tilde{V}\Big)^{ 2}-2\tilde{V}\frac{\rho_{r}}{\rho_{\ast}}
+\frac{\tilde{c}}{a^{4}}}\,\Bigg]^{2} -36\Bigg{\}}.
\label{H1bb}
\end{equation}
The branch $\epsilon=+1$ is completely new comparing to the standard
braneworld models since the scale factor is bounded from above for any
value of $\tilde{V}$. Therefore, contrary to the branch $\epsilon=-1$,
here, there is no pure late-times linearization regime. However, expanding
the expression (\ref{H1bb}), there is a term linear in $\rho_{m},\rho_{r}$, so
Newton's constant $G_{N}$ can also here be identified. More precisely it is
$H^{2}+\frac{k}{a^{2}}
-\frac{\mathcal{C}}{a^{4}}=\gamma(\rho_{m}+\frac{\rho_{r}}{2})+...$, where ...
do not contain terms linear in $\rho_{m},\rho_{r}$, and $\gamma=\frac{V}{144M^{6}}$.
Therefore, associating $G_{N}$ with $\rho_{m}$ we have the identification
\begin{equation}
\gamma=\frac{V}{144M^{6}}\equiv\frac{8\pi G_{N}}{3}\,.\\
\label{gamma1v}
\end{equation}
Going back to equation (\ref{H1bb}), we eliminate the parameter $M$ and we rewrite the
expansion rate for $\epsilon=+1$ as
\begin{eqnarray}
H^{2}+\frac{k}{a^{2}}-\frac{\mathcal{C}}{a^{4}}=\frac{4\pi G_{N}\rho_{\ast}}{3\tilde{V}}
\Bigg[&&\!\!\!\tilde{V}\frac{2\rho_{m}+\rho_{r}}{\rho_{\ast}}+\Big(\frac{\rho_{m}
+\rho_{r}}{\rho_{\ast}}\Big)^{2}
+\tilde{V}^{2}-18+\frac{\tilde{c}}{2a^{4}}\nn\\
&&\!\!-\Big(\frac{\rho_{m}\!+\!\rho_{r}}{\rho_{\ast}}\!+\!\tilde{V}\Big)
\sqrt{\Big(\frac{\rho_{m}\!+\!\rho_{r}}{\rho_{\ast}}\!+\!\tilde{V}\Big)^{2}
\!-\!2\tilde{V}\frac{\rho_{r}}{\rho_{\ast}}
\!+\!\frac{\tilde{c}}{a^{4}}}\,\Bigg]\,.
\label{Hplus1b}
\end{eqnarray}
This expression takes the standard form
\begin{equation}
H^{2}+\frac{k}{a^{2}}-\frac{\mathcal{C}}{a^{4}}= \frac{8\pi G_N}{3}
(\rho_m+\rho_r+\rho_{DE}),
\label{Hminus1cc}
\end{equation}
where
\begin{equation}
\rho_{DE}\!=\!\frac{ \rho_{\ast}}{2\tilde{V}}
\Bigg[\!\Big(\frac{\rho_{m}\!+\!\rho_{r}}{\rho_{\ast}}\Big)^{\!2}
\!-\frac{\tilde{V}\rho_{r}}{\rho_{\ast}}+\tilde{V}^{2}-18+\frac{\tilde{c}}{2a^{4}}
-\Big(\frac{\rho_{m}\!+\!\rho_{r}}{\rho_{\ast}}+\tilde{V}\Big)
\sqrt{\Big(\!\frac{\rho_{m}\!+\!\rho_{r}}{\rho_{\ast}}\!+\!\tilde{V}\!\Big)^{2}
\!\!-\!2\tilde{V}\frac{\rho_{r}}{\rho_{\ast}}
\!+\!\frac{\tilde{c}}{a^{4}}}\,\Bigg].
\label{rhoDE}
\end{equation}
Defining the density parameters as in (\ref{Omegam})-(\ref{OmegaC}), we find equation
(\ref{Hminus1dd}), where $\rho_{DE}(z)$ is now given by
\begin{eqnarray}
&&\!\!\!\!\!\!
\rho_{DE}(z)=\frac{\rho_{\ast}}{2\tilde{V}}\Bigg\{\left(\frac{3H_{0}^{2}\Omega_{m0}}{8\pi G_{N}\rho_{\ast}}
(1+z)^{3}+\frac{3H_{0}^{2}\Omega_{r0}}{8\pi G_{N}\rho_{\ast}}
(1+z)^{4}\right)^{2}-\frac{3H_{0}^{2}\Omega_{r0}\tilde{V}}{8\pi G_{N}\rho_{\ast}}
(1+z)^{4}+\tilde{V}^{2}-18\nn\\
&&\quad\quad\quad\quad\quad\quad
+\frac{\tilde{c}}{2}(1+z)^{4}-\left(\frac{3H_{0}^{2}\Omega_{m0}}{8\pi G_{N}\rho_{\ast}}
(1+z)^{3}+\frac{3H_{0}^{2}\Omega_{r0}}{8\pi G_{N}\rho_{\ast}}
(1+z)^{4}+\tilde{V}\right)\mathcal{A}(z)\Bigg\},
\label{poulia}
\end{eqnarray}
\begin{figure}[!]
\begin{center}
\includegraphics[width=16cm]{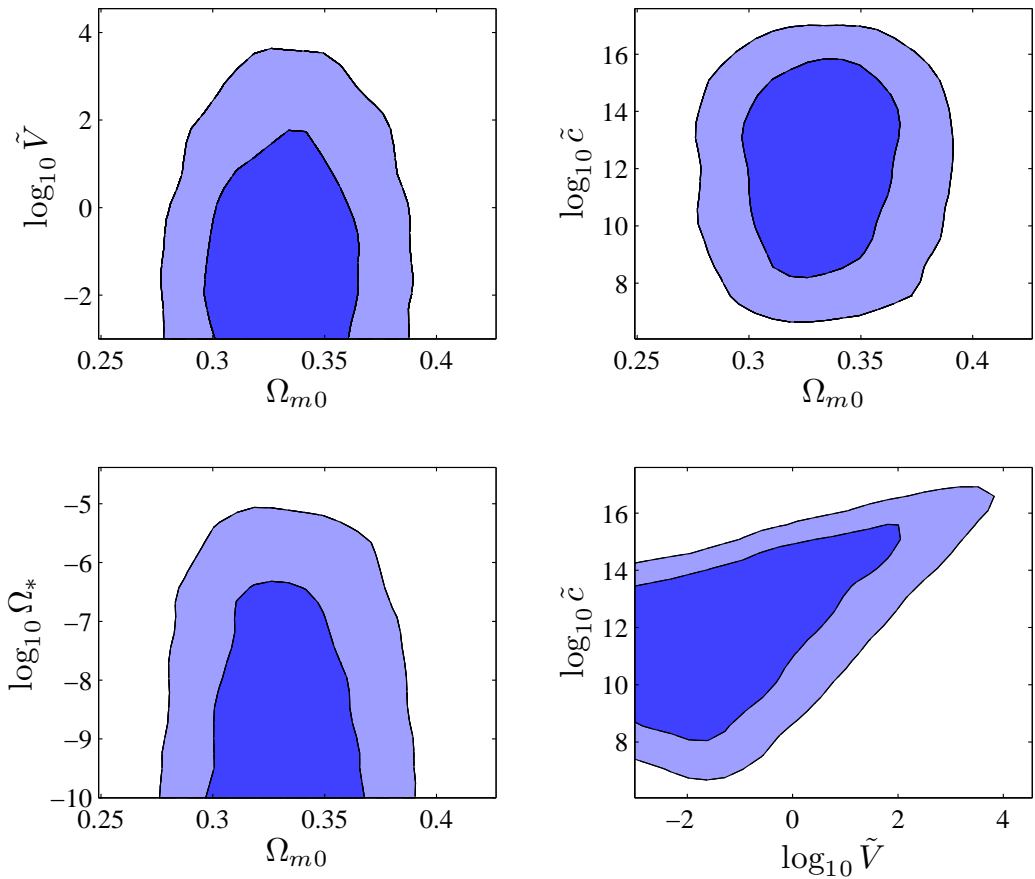}
\caption{(Color Online) {\it{Two-dimensional likelihood contours in the
($\Omega_{m0},\tilde{V}$), ($\Omega_{m0},\log_{10}\tilde{c}$),
($\Omega_{m0}, \Omega_\ast$) and ($\tilde{V}, \log_{10}\tilde{c}$) planes,
for the $\epsilon=+1$ branch, from the SnIa+BAO data
combinations. The lighter regions correspond
to 2$\sigma$ confidence level, while the darker region  correspond to
1$\sigma$ confidence level.}}} \label{positive}
\end{center}
\end{figure}
with
\begin{equation}
\mathcal{A}(z)=\sqrt{
\left(\frac{3H_0^2\Omega_{m0}}{8\pi
G_N \rho_{\ast}}(1\!+\!z)^3+\frac{3H_0^2\Omega_{r0}}{8\pi G_N \rho_{\ast}}(1\!+\!z)^4
+\tilde{V}\right)^{\!2}
-\frac{3H_0^2\Omega_{r0}\tilde{V}}{4\pi G_N \rho_{\ast}}(1\!+\!z)^4+\tilde{c}(1\!+\!z)^4}\,.
\end{equation}

In summary, Eq. (\ref{Hminus1cc}) is the one we will fit, with
$\mathcal{C}$,
$\tilde{c}$, $\tilde{V}$, $\rho_{\ast}$ and $\Omega_{m0}$ as parameters (again for
simplicity we fix
$\Omega_{k0}$ to its (Planck+WP+highL+BAO) best fit values, namely
$\Omega_{k0}=-0.0003$
\cite{planckfit}). Additionally, we include the direct $H_0$ probe from
the Hubble Space Telescope (HST) observations of Cepheid variables with
$H_0=73.8\pm2.4$ ${\rm km\,s^{-1}\,Mpc^{-1}}$.
Similarly to the
previous subsection, we can safely neglect $\mathcal{C}$ since it is
negligible according to BBN analysis. Finally, instead of $\rho_{\ast}$ it
proves more convenient to introduce the dimensionless quantity
(\ref{Omegastar}), namely
$\Omega_\ast \equiv\frac{\rho_\ast}{\rho_{c0}}$,
where
$\rho_{c0}$ is the present critical energy density of the  Universe.

We use combined SnIa and BAO data to constrain $\tilde{c}$, $\tilde{V}$,
$\Omega_\ast$ and $\Omega_{m0}$.
In Fig. \ref{positive} we present the corresponding two-dimensional
likelihood contours. Firstly, note that in this case $\tilde{V}$ is not
theoretically restricted to values greater than 3 and in particular it is
constrained in much smaller values, namely  $\log_{10}\tilde{V} < 2.0$ (95\%
C.L. upper limit). Additionally, note that
since at late times $\rho_{DE}$ acquires negative values,  the
constraint on $\Omega_\ast$ is very close to zero, namely
$\log_{10}\Omega_\ast < -5.5$ (95\% C.L.). Due to the strong degeneracy
between $\Omega_\ast$ and  $\tilde{c}$,  the constraints on    $\tilde{c}$
are very different from those in the $\epsilon=-1$ branch
case, namely $7.7 <
\log_{10}\tilde{c} < 15.9$ (95\% C.L.). However, note that the
minimal $\chi^2$ for this case is  $\chi^2\approx688$, that is much higher
than that for the $\epsilon=-1$ branch case, which means that the $\epsilon=+1$
branch case is not favored by observations. This can be additionally seen by
calculating the corresponding  age of the universe, which is much smaller than
the $\Lambda$CDM value. However, although this branch is
not favored by late-times observations, due to that $H^{2}\approx \text{const.}$
at early times, it could still play an important role in the inflationary regime.

\section{Conclusions}
\label{Conclusions}

In this work we constrained an alternative 5-dimensional braneworld cosmology
using observational data.
The difference with the standard braneworld cosmology refers to the
adaptation of alternative matching
conditions introduced in \cite{kof-ira} which generalize the conventional
matching conditions. The reasons
for this consideration are possible theoretical deficiencies of the standard
junction conditions, namely
the need for consistency of the various codimension defects and the existence
of a meaningful equation
of motion at the probe limit. Instead of varying the brane-bulk action with
respect to the bulk
metric at the brane position and derive the standard matching conditions, we
vary with respect to the brane
embedding fields in a way that takes into account the gravitational
back-reaction of the brane onto the bulk.

The proposed gravitating Nambu-Goto matching conditions may be close to the
correct direction of finding realistic
matching conditions since they always have the Nambu-Goto probe limit
(independently of the gravity theory, the
dimensionality of spacetime or codimensionality of the brane), and moreover,
with these matching conditions,
defects of any codimension seem to be consistent for any (second order)
gravity theory. Compared to the conventional
5-dimensional braneworld cosmology, the new one possesses an extra
integration constant, which if set to zero  reduces the new cosmology to the
conventional braneworld one.

In the present work we extended the codimension-1 cosmology of
\cite{kof-zar} by
allowing both a matter and a
radiation sector in order to extract observational constraints on the
involved model parameters. In particular, we used data  from supernovae type
Ia (SNIa) and Baryon Acoustic Oscillations (BAO), along with arguments from
Big Bang Nucleosynthesis (BBN) in order to
construct the corresponding
probability contour-plots for the parameters of the theory.

Concerning the first ($\epsilon=-1$) branch of cosmology, we found that the
parameters $\tilde{V}$ and $\tilde{c}$ that quantify the deviation from the
Randall-Sundrum scenario, are constrained very close to their RS values as
expected. However, a departure from Randall-Sundrum scenario is still
allowed, and moreover, the corresponding $\chi^2$ is the same for both models.
This means that braneworld models with gravitating Nambu-Goto matching
condition are in ``equal'' agreement with observations with standard
braneworld cosmology. However, application of the AIC criterion shows that
both standard braneworld cosmology, as well as the extended scenario of the
present work,
are less favored by the data if we compare them with the concordance
$\Lambda$CDM cosmology since the two extra parameters do not improve the
fitting behavior.  Furthermore, the
obtained age of the universe is $12.23\ \text{Gyr}\leq
t_0\leq14.13\ \text{Gyr}$, which is an additional observational advantage of
the model. Finally, concerning the fundamental mass scale $M$, the
current age estimations imply that the preferred values of  $M$ lie
well below the GeV scale.

Concerning the second ($\epsilon=+1$) cosmological branch, which is
completely new and with no correspondence in Randall-Sundrum scenario, we
extracted the corresponding likelihood contours. Although this case is still
compatible with observations, the corresponding minimal $\chi^2$ is much
higher than that for the $\epsilon=-1$ branch case, which means that this
branch case is not favored by late-times observations. However, although this branch is
not favored by late-times observations, due to that $H^{2}\approx \text{const.}$
at early times, it could still play an important role in the inflationary regime.

In summary, cosmology with gravitating Nambu-Goto matching conditions offers an
extension to the standard Randall-Sundrum scenario. Apart from interesting
solutions, we see that it is in agreement with observations since the data
allow for a small deviation from Randall-Sundrum cosmology. Therefore, it
should be worthy to further study the cosmological implications of the
model, such as the inflationary behavior and the late-times asymptotic
features, since especially a successful  inflationary regime is something
that cannot be obtained in the framework of $\Lambda$CDM cosmology.

\begin{acknowledgments}
The research of ENS is implemented within the framework of the Action
``Supporting Postdoctoral Researchers'' of the Operational Program
``Education and Lifelong Learning'' (Actions Beneficiary: General Secretariat
for Research and Technology), and is co-financed by the European Social Fund
(ESF) and the Greek State. J.-Q. X. is supported by the National Youth 
Thousand Talents Program, the National Science 
Foundation of China under Grant No. 11422323, and 
the Strategic Priority Research Program “The Emergence 
  of Cosmological Structures” of the Chinese Academy of 
    Sciences, Grant No. XDB09000000. 
\end{acknowledgments}

\begin{appendix}

\section{Observational data and constraints}
\label{Observational data and constraints}

In this Appendix we review the main procedures of observational fittings used
in the present work, namely Type Ia Supernovae (SNIa) and Baryon Acoustic
Oscillations (BAO).\\

{\it{a. Type Ia Supernovae constraints}}\\

We use the Union 2.1 compilation of SnIa data \cite{Suzuki:2011hu} in order
to incorporate Supernovae type Ia constraints. This is a heterogeneous
data set, which includes data from the Supernova Legacy Survey, the
Essence survey and the Hubble-Space-Telescope observed distant
supernovae.

The $\chi^2$ for this analysis is written as
\begin{equation} \chi ^2 _{SN} =
\frac{{\sum\limits_{i = 1}^N {\left[ {\mu _{\text{obs} } \left(
{z_i } \right) - \mu _{\rm th} \left( {z_i } \right)} \right]} ^2
}}{{\sigma^{2} _{\mu,i} }},
\end{equation}
where $N=580$ is the number of SNIa data points. In the above expression
$\mu_{\rm obs}$ is the observed distance modulus, which is defined as the difference of the
supernova apparent magnitude from its absolute one.
Furthermore, $\sigma_{\mu,i}$ are the errors in
the observed distance moduli, which are assumed to be uncorrelated and
Gaussian, arising from a variety of sources. If we introduce the usual
(dimensionless) luminosity distance $D_{L}(z;a_i)$,
calculated by
\begin{equation}
D_{L}\left(z;a_i\right)\equiv\left(1+z\right)
\int^{z}_{0}dz'\frac{H_0}{H\left(z';a_i\right)},
\end{equation}
with $H_0$ the present Hubble parameter, then the theoretical
distance modulus $\mu_{\rm th}$  has a dependence on the model parameters
$a_i$  as
\begin{equation}  \mu_{\rm
th}\left(z\right)=42.38-5\log_{10}h+5\log_{10}\left[D_{L}
\left(z;a_i\right)\right].
\end{equation}
Finally, the marginalization over the present Hubble parameter is performed
following \cite{perivol1}, which eventually provides the $\chi^2$ likelihood
contours for the model parameters that are involved.
\\

{\it{b. Baryon Acoustic Oscillation constraints}}\\

In order to handle the baryon acoustic oscillation (BAO) observational
constraints we use the definition \cite{oldbao}
\begin{eqnarray}
A\equiv D_{V}(z=0.35)\frac{\sqrt{\Omega_m H^2_0}}{0.35c}=0.469\pm0.017~,
\end{eqnarray}
where $c$ is the light speed. In the above expression we have defined the
``volume distance''
$D_{V}\left(z\right)$ as
\begin{equation}
D_{V}\left(z\right)\equiv\left[\frac{\left(1+z\right)^2 D_{A}^{2}(z) z
}{H(z)}\right]^{1/3},
\end{equation}
where
\begin{equation}
D_{A}\equiv r\left(z\right) /\left(1+z\right)
\end{equation}
is the angular diameter distance. Finally, the BAO likelihood is written as
\begin{equation}
\chi^2_{BAO}= \frac{(A-0.469)^2}{0.017^2}~.
\end{equation}

\end{appendix}

\end{document}